# Optical vortex Brillouin laser

Xinglin Zeng[1,*], Philip St.J. Russell[1], Yang Chen[1], Zheqi Wang[1], Gordon K. L. Wong[1], Paul Roth[1], Michael H. Frosz[1], and Birgit Stiller[1,2]

[1]Max Planck Institute for the Science of Light, Staudtstr. 2, 91058 Erlangen, Germany
[2]Department of Physics, Friedrich-Alexander University, Staudtstr. 2, 91058 Erlangen, Germany

[*]Corresponding author: xinglin.zeng@mpl.mpg.de

**Optical vortices, which have been extensively studied over the last decades, offer an additional degree of freedom useful in many applications, such as optical tweezers and quantum control. Stimulated Brillouin scattering, providing a narrow linewidth and a strong nonlinear response, has been used to realise quasi-continuous wave (CW) lasers. Here, we report stable oscillation of optical vortices and acoustic modes in a Brillouin laser based on chiral photonic crystal fibre, which robustly supports helical Bloch modes (HBMs) that carry circularly-polarized optical vortex and display circular birefringence. We implement a narrow-linewidth Brillouin fibre laser that stably emits 1st- and 2nd-order vortex-carrying HBMs. Angular momentum conservation selection rules dictate that pump and backward Brillouin signals have opposite topological charge and spin. Additionally, we show that when the chiral PCF is placed within a laser ring cavity, the linewidth-narrowing associated with lasing permits the peak of the Brillouin gain that corresponds to acoustic mode to be measured with resolution of 10 kHz and accuracy of 520 kHz. The results pave the way to a new generation of vortex-carrying SBS systems with applications in quantum information processing, vortex-carrying nonreciprocal systems.**

## 1. Introduction

Stimulated Brillouin scattering (SBS) in optical fibres, in which guided light is parametrically reflected by coherent acoustic phonons, provides a powerful and flexible mechanism for controlling light. Since its first demonstration [1], SBS has been explored and exploited in many different systems, especially optical fibres [2] and integrated photonics [3]. Among the many applications of SBS, Brillouin laser oscillators have been used in mode-locked lasers [4], microwave oscillators [5] and optical gyroscopes [6].

Light fields carrying optical vortices have many potential applications, for example optical tweezers [7] and classical or quantum communications [8,9]. Brillouin lasing with optical vortices has attracted interest [10,11], since Brillouin scattering provides a narrow linewidth and a nonlinear response that is typically stronger than Kerr and Raman interactions in most transparent media. However, a major challenge in current SBS-related vortex emission schemes is maintaining the vorticity as the light circulates in the cavity; the mode controllers that are typically used require constant adjustment and introduce optical loss. The recent emergence of chiral photonic crystal fibres (PCF) [12] − a unique platform for studying the behavior of light in chiral structures that are infinitely extended in the direction of the twist − has enabled investigation of SBS process in the presence of chirality [13].

Here we report a Brillouin fibre laser in which chiral photonic crystal fibre (PCF) is used to robustly preserve circularly polarized vortex modes over long distances, permitting topology-selective Brillouin gain [14]. In particular, angular momentum conservation dictates that the topological charge and spin of the backward Brillouin signal are opposite in sign to those of the pump. Making use of this fact, we experimentally demonstrate stable Brillouin lasing in different-order circularly polarized vortex modes, supported by acoustic modes guided in the chiral PCF core. Linewidth-narrowing in the laser cavity allows the frequency shift of each individual Brillouin gain peak to be measured with resolution of 10 kHz and accuracy of 520 kHz, corresponding to the laser linewidth and cavity free spectral range, respectively. The results are consistent with theoretical predictions. Vortex carrying Brillouin lasers have potential applications in narrow-linewidth sources for quantum information processing, twisted photon-phonon studies and optical tweezers.

## 2. Results and Discussion

Two types of chiral multicore PCF were used in the measurements: $C_3$ PCF has 3-fold rotational symmetry with a twist pitch of 5 mm [15] (inset in Figure 1a) and $C_6$ PCF has 6-fold rotational symmetry with a twist pitch of 7.2 mm (inset in Figure 1b). Both chiral PCFs were formed by replacing specific preform capillaries with solid glass rods and drawing the fibres from a spinning preform. Chiral PCFs with $N$-fold rotationally symmetric (symmetry class $C_N$) support circularly-polarized helical Bloch modes (HBMs) with $m$-th order azimuthal harmonics that carry optical vortices with azimuthal order $\ell_A^{(m)} = \ell_A^{(0)} + Nm$ ($\ell_A^{(0)}$ is the principal order). The $\ell_A^{(m)}$ is obtained in cylindrical coordinates and for pure circularly polarized fields can be related to the topological charge $\ell_T^{(m)}$ in Cartesian coordinates via $\ell_A^{(m)} = \ell_T^{(m)} + s$ (spin $s = +1$ for left-circular polarization) simply by rotating the reference frame. The shorthand $[\ell_T, s]$ will be used to denote each circularly polarized vortex-carrying HBM, where $\ell_T$ is defined to be $\ell_T^{(0)}$. Figure 1a and 1b show the modal refractive indices (blue curves) in both $C_3$ and $C_6$ PCFs against topological charge $\ell_T$, estimated from the analytical model [16]. The red circles are numerically calculated modal indices of each $\ell_T$. Both $[\ell_T, +1]$ and $[\ell_T, -1]$ collapse on to one curve, as the modes with opposite spin are only very weakly birefringent. The calculated mode, phase and polarization states distributions of [+1, −1], [−1, +1], [+2, −1] and [−2, +1] modes are shown in Figure 1. The index difference is ~$8\times10^{-4}$ between [+1, ±1] and [−1, ±1] modes and ~$7\times10^{-4}$ between [+2, ±1] and [−2, ±1] modes. The measured loss of both [±1, ±1] and [±2, ±1] modes are 0.017 dB/m and 0.04 dB/m, respectively.

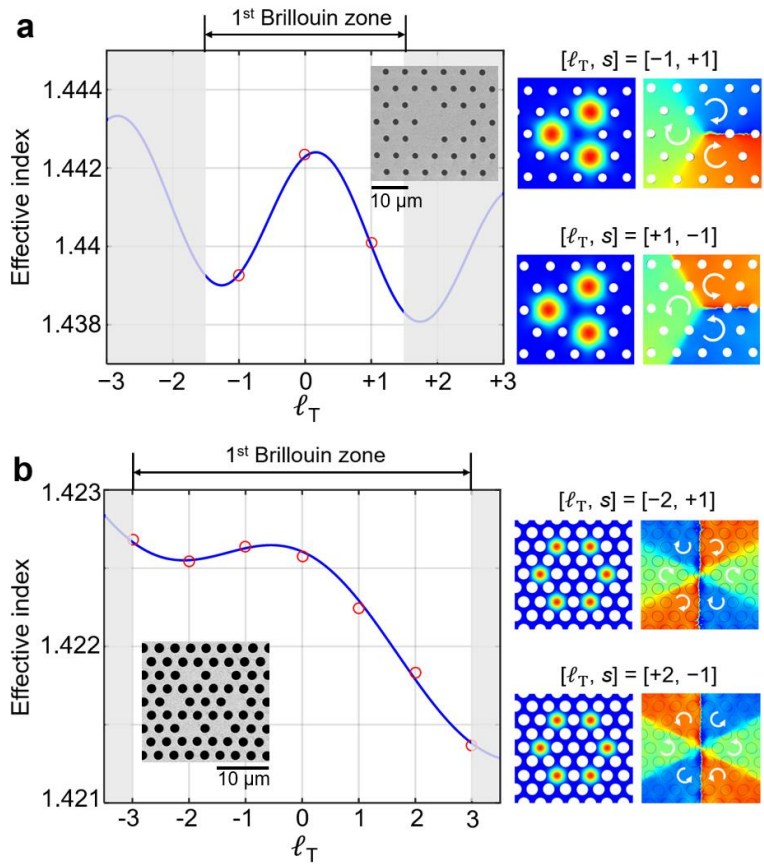

**Figure 1.** a) Calculated effective refractive index versus topological charge $\ell_T$ in $C_3$ PCF (left) and mode, phase and polarization states distributions (right) of $[+1, -1]$ and $[-1, +1]$ modes. The inset shows a scanning electron micrograph (SEM) of $C_3$ PCF. b) Same as a), for PCF $C_6$.

Details of the Brillouin measurement setup are available in supporting information S1. Figure 2a shows the spontaneous Brillouin spectra measured after injecting a 1 W [−1, +1]$_P$ mode into a 200 m length of $C_3$ PCF and a 1 W [−2, +1]$_P$ mode into a 200 m length of $C_6$ PCF (subscript P denotes the pump mode). The Brillouin peak frequencies are 11.054 GHz for the [−1, +1]$_P$ mode and 10.994 GHz for the [−2, +1]$_P$ mode. The spontaneous Brillouin spectrum for [−1, +1]$_P$ pumping is slightly asymmetric, deviating from a perfect Lorentzian, which suggests that more than one acoustic mode is excited. Although it is impossible to resolve these closely spaced peaks by normal heterodyning techniques, they could be precisely measured when the linewidths of Stokes signals are narrowed during light circulation in the laser ring cavity (the details will be discussed later in this paper). Figure 2b shows the Brillouin gain spectra measured in 200 m lengths of PCF $C_3$ (blue) and PCF $C_6$ (red) using a specially designed pump-seed setup (see supporting information S2). As dictated by angular momentum conservation [14], the Brillouin gain is only significant when pump and seed have opposite spin and topological charge, reaching peak values of 0.022 $W^{-1}m^{-1}$ for [−1, +1]$_P$/[+1, −1]$_S$ pump/seed and 0.19 $W^{-1}m^{-1}$ for [−2, +1]$_P$/[+2, −1]$_S$, where subscript S denotes Stokes wave.

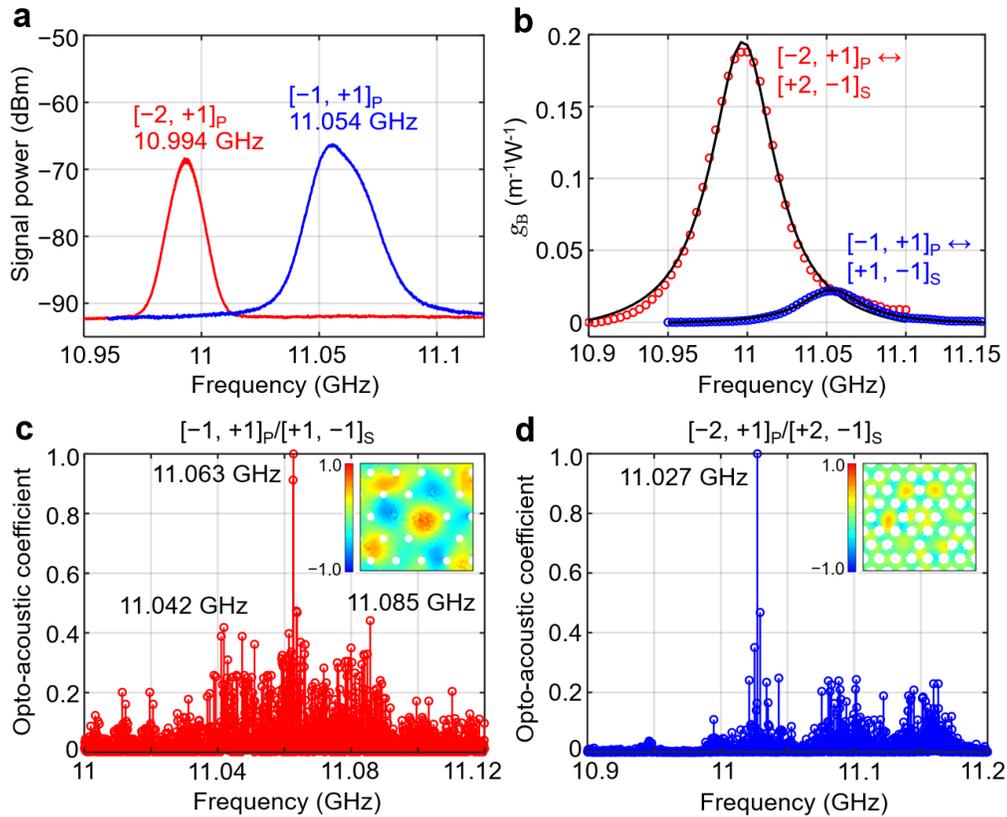

**Figure 2.** a) Spontaneous Brillouin spectra generated by 1550 nm pumping $C_3$ PCF with a 1 W [−1, +1]$_P$ mode (blue) and $C_6$ PCF with a 1 W [−2, +1]$_P$ mode (red). b) Measured Brillouin gain spectra for pumping the [−1, +1]$_P$ mode in $C_3$ PCF (blue) and the [−2, +1]$_P$ mode in $C_6$ PCF (red). The black lines are Lorentzian fits. c) Numerically calculated opto-acoustic coupling coefficients for each acoustic mode that satisfies phase-matching (see text), normalized to the highest value of κ. d) Same as c), for $C_6$ PCF. The insets in c) and d) show the axial displacement (normalized to the square-root of the power) for the peaks at 11.063 GHz and 11.027 GHz.

To confirm the experimental results, we used COMSOL software to calculate the opto-acoustic coupling coefficient [17]:

$$\kappa \propto \int\int dxdy E_P E_S^* p_{ijkl}\varepsilon_{kl}, \qquad (1)$$

where $E_P$ and $E_S$ are the scalar electric field distributions of circularly polarized vortex-carrying pump and Stokes modes, $p_{ijkl}$ is the elasto-optic tensor and $\varepsilon_{kl}$ is the strain tensor associated with the acoustic mode. The material parameters, such as Young's modulus, Poisson's ratio and coefficient of thermal expansion, are set to the default values for fused silica in the COMSOL materials library. The acoustic wavevector is first set to $\Delta\beta = 2\pi(n_P+|n_S|)/\lambda_P$, and the frequencies of the acoustic modes (typically ~500) that share this wavevector are calculated. Equation (1) is then evaluated for each of these modes, and the results normalised to the mode with the highest value of κ. Figure 2c shows the results for [−1, +1]P/[ +1, −1]S modes in the $C_3$ PCF, revealing that, although a large number of acoustic modes satisfy phase-matching within the frequency range from 11 GHz to 11.12 GHz, only the modes at 11.063 GHz (corresponding to the experimental peak at 11.054 GHz) have significant overlap with the optical modes and thus are dominant. The inset shows power-normalized axial displacements of the acoustic mode at 11.063 GHz. However, it is possible that the calculated parasitic acoustic modes at 11.042 GHz and 11.085 GHz are weakly excited in the experiment, causing the Brillouin spectrum to deviate from a perfect Lorentzian. Figure 2d shows the same calculations as Figure 2c, but for [−2, +1]P/[ +2, −1]S modes in $C_6$ PCF in the frequency range from 10.9 GHz to 11.2 GHz. Although three peaks are obtained from the simulations, only the main peak at 11.027 GHz was experimentally observed (corresponding to the peak at 10.994 GHz). We attribute this to much weaker opto-acoustic coefficients, resulting in the other two peaks being below the noise floor.

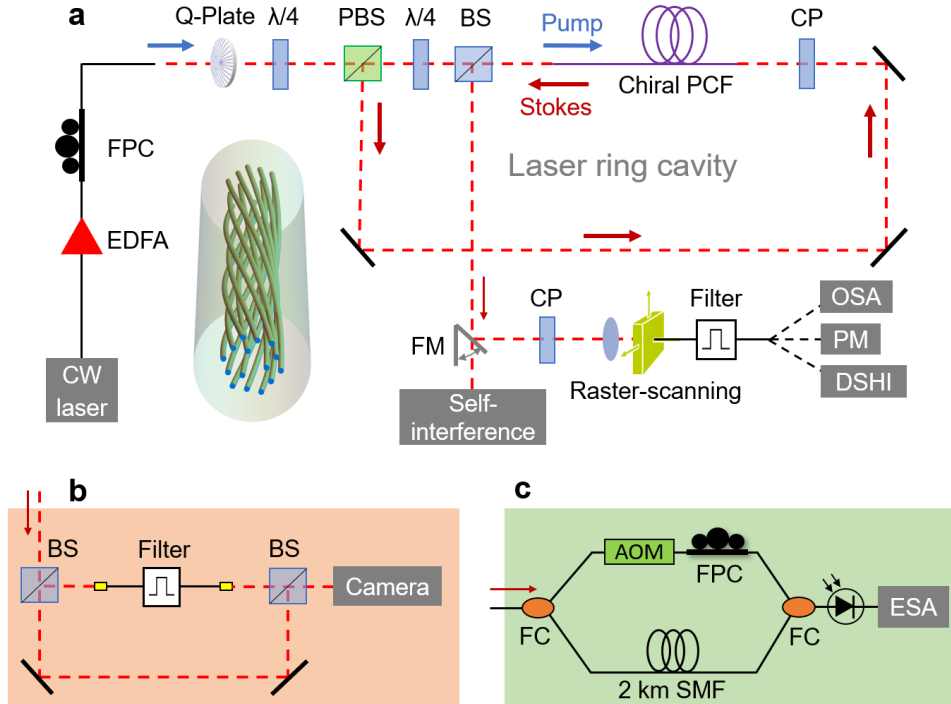

**Figure 3.** a) Experimental setup of vortex Brillouin laser. FPC: polarization controller, CP: circular polarizer, BS: beam splitter, PBS: polarizing beam splitter, FM: flip mirror, OSA: optical spectrum analyzer, PM: power meter, DSHI: delayed self-heterodyne interference. Inset: 3D sketch of the chiral PCF. b) Experimental setup of spatial self-interference for Stokes topological charge measurement. c) Experimental setup of delayed self-heterodyne interference for laser linewidth measurement. AOM: acousto-optical modulator, FC: fibre coupler, SMF: single mode fibre.

Using 200 m lengths of $C_3$ and $C_6$ PCF we constructed two Brillouin fibre lasers generating different-order circularly polarized-vortex modes. The experimental setup is shown in Figure 3a. The CW pump light passes sequentially through an erbium-doped fibre amplifier (EDFA), a polarization controller (FPC), a Q-plate, a λ/4 plate, a polarizing beam-splitter (PBS) and a second λ/4 plate, so as to launch a circularly-polarized vortex mode into the laser. Since the Stokes signal is orthogonally polarized relative to the pump, it is reflected by the PBS and thus circulates inside the cavity. A circular polarizer (a combination of a λ/4 plate and a polarizer) blocks the transmitted pump light, while letting the backward Stokes signal propagate freely. A beam splitter (BS) was used to couple 10% of the laser signal out of the cavity and another circular polarizer enabled measurement of the Stokes polarization state. In order to precisely measure the mode profile at the laser output, we designed a near-field scanning Brillouin analyzer (NBA) that includes a fibre raster-scanning stage, a narrow-band (6 GHz) filter and equipment for signal analysis (optical spectrum analyser (OSA), power meter, delayed self-heterodyne interference system). The collected laser light, which contains small amounts of pump light (caused by Fresnel reflections or Rayleigh scattering), is collected pixel by pixel using a fibre raster scanning stage. The signal is then filtered to remove the pump frequency and then analysed. An additional flip mirror was used to steer the laser beam into a self-interference setup (Figure 3b) for topological charge measurement. In the setup, the Stokes signal was split in two, one half being spatially filtered in a single-mode fibre to produce a divergent near-Gaussian beam which was then superimposed on the other half, resulting in spiral patterns of fringes related to the topological charge. These patterns were imaged using a CCD camera after filtering out any stray pump light with a narrow-band filter.

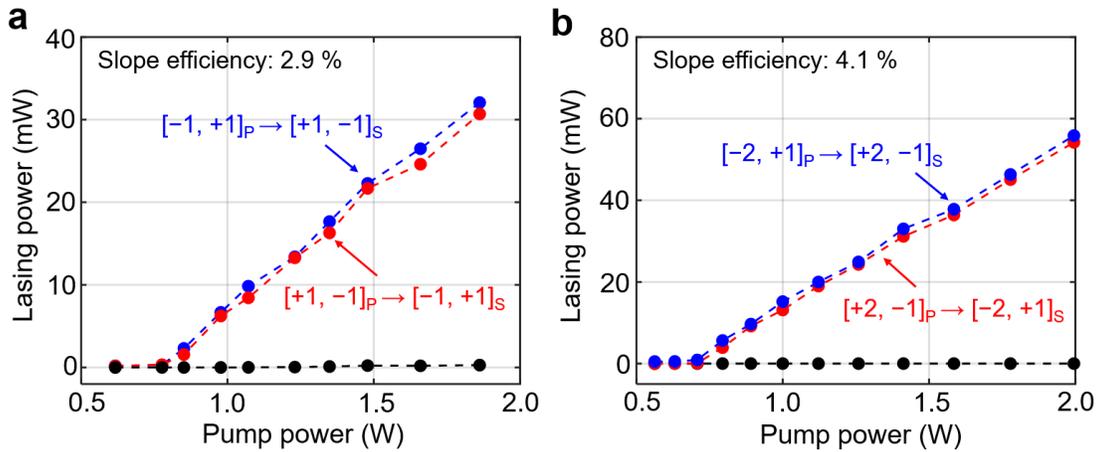

**Figure 4.** a) Output power of $[+1, -1]_S$ and $[-1, +1]_S$ Brillouin lasers in $C_3$ PCF, as a function of $[-1, +1]_P$ and $[+1, -1]_P$ power. b) Output power of $[+2, -1]_S$ and $[-2, +1]_S$ Brillouin lasers in $C_6$ PCF, as a function of $[-2, +1]_P$ and $[+2, -1]_P$ power.

The laser output power is plotted against $[\pm1, \pm1]_P$ power in Figure 4a and $[\pm2, \pm1]_P$ power in Figure 4b. The threshold power, defined as the point at which the Stokes power equals 1% of the pump power, is 800 mW for $[\pm1, \pm1]_P$ and 700 mW $[\pm2, \pm1]_P$, with slope efficiencies of ~2.9% and ~4.1%. Lasing commences when the Brillouin gain exceeds the laser cavity round-trip loss of 9.4 dB for PCF $C_3$ and 12 dB for PCF $C_6$. The cavity length was adjusted so that a single cavity mode coincided with the maximum Brillouin gain. The cavity included 200 m of PCF and 2.2 m of free space, resulting in a free spectral range of ~520 kHz, which given the 45 MHz FWHM Brillouin gain linewidth means that 86 cavity longitudinal laser modes can participate in lasing. However, in a situation with strong mode competition [18], only one mode survives and oscillates in the laser cavity. Note that both the threshold power and slope efficiency for $[+\ell_T, s]_P$ and $[-\ell_T, s]_P$ pumping are almost the same. This is because an identical acoustic mode is excited in each case. Nevertheless, some tiny differences are still observed in the two cases and we attribute this to slight circular dichroism in propagation loss. The laser output power was stable, with fluctuations of <1% over 1 hour. Mode hopping (within the 1 MHz range) nevertheless constantly occurs, since the lasing cavity is not isolated from the laboratory environment.

The spectrum measured by the high-resolution OSA just before the filter confirmed the presence of a $[+1, -1]_S$ laser signal with frequency 11.054 GHz below the pump frequency (Figure 5a). The upper panels of Figure 5b show the near-field profiles of $[-1, +1]_P$ and $[-2, +1]_P$ pump light measured by the camera, together with the spiral fringe patterns measured by interference with a divergent Gaussian beam. The lower panel of Figure 5b shows the near-field profiles of the corresponding laser signals measured by the NBA system, along with their spiral fringe patterns. Again, it is seen that pump and laser signals have opposite topological charge and spin. Since grating-based OSAs and Fabry-Perot interferometers typically have insufficient spectral resolution (a few GHz and tens of MHz respectively) [19], we implemented a "sub-coherence" delayed self-heterodyne interference (DSHI) [20], which is capable to measure sub-kHz linewidths. Figure 3c shows the measurement setup. The Brillouin laser output is launched into an SMF and split into two paths at a 3 dB fibre coupler. One path is frequency-shifted by 200 MHz using an acousto-optic modulator (AOM), and the other path is transmitted through a 2 km length of SMF, so that it becomes sub-correlated from the frequency-shifted light. Figure 5c and Figure 5d show the measured spectra of $[+1, -1]$ and $[+2, -1]$ vortex Brillouin lasers. The sharp side-peaks close to the center are artefacts from mechanical vibrations in the pump laser. Linewidths of 10 kHz for $[+1, -1]$ Brillouin laser and 8 kHz for $[+2, -1]$ Brillouin laser were obtained by fitting the measured laser spectra to the sub-coherence lineshape function. A comparison between lasing and spontaneous Stokes spectra (generated without optical cavity feedback) shows a line-narrowing factor of ~4×10³ for the $[+1, -1]_S$ laser and ~5×10³ for the $[+2, -1]_S$ laser (Figures 5c and 5d).

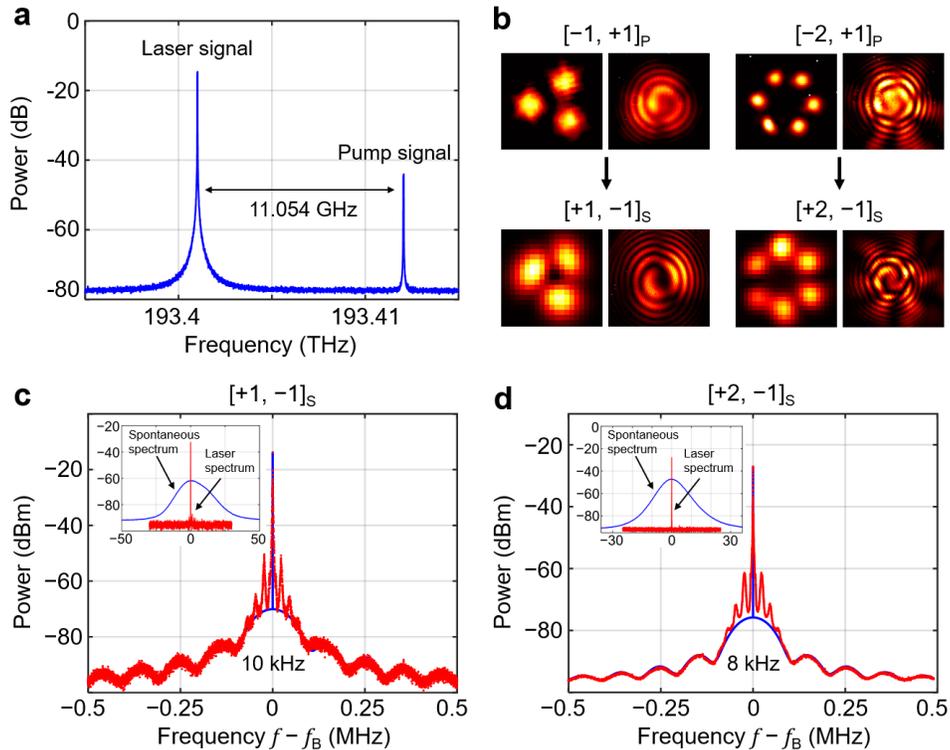

**Figure 5.** a) Power spectrum of $[+1, -1]_S$ laser output before the narrow-band filter, measured by the OSA. b) Measured mode profiles and interferometric spiral fringe patterns of $[-1, +1]_P$ (or $[-2, +1]_P$) pump and $[+1, -1]_S$ (or $[+2, -1]_S$) vortex Brillouin laser signals; c) The laser spectrum (red) for the $[+1, -1]_S$ mode in $C_3$ PCF, measured by sub-coherence delayed self-heterodyning and compared with a theoretical fit (blue). The inset compares the spontaneous Stokes spectrum from $C_3$ PCF with the line-narrowed intracavity laser spectrum measured by delayed self-heterodyne system (the axis labels are same as those of main figure, $f_B$ is the Brillouin frequency). d) Same as c), for the $[+2, -1]_S$ lasing mode in $C_6$ PCF.

As discussed in the Figure 2a, the spontaneous Brillouin spectra for $[\pm 1, \pm 1]_P$ pump modes in $C_3$ PCF are slightly asymmetric, deviating from perfect Lorentzian lineshapes. This can be attributed to weak excitation of other acoustic modes, causing Brillouin shifts only ~20 MHz away from the main peak. Although these closely-spaced peaks are difficult to be resolved experimentally by spontaneous scattering or pump-probe techniques, precise measurements can be made using the line-narrowing associated with Brillouin lasing. Figure 6a shows the spectrum obtained by normal heterodyning between the laser signal and a local oscillator at the pump frequency. Three Brillouin lasing peaks are seen, at 11.04 GHz, 11.054 GHz and 11.068 GHz (blue curve), the one at 11.054 GHz being the strongest. The experimental results are in excellent agreement with the simulations in Figure 2c. The insets show the normalized axial displacement of the three associated acoustic modes. Linewidth-narrowing to ~10 kHz permits the centre-band frequencies of the three peaks to be easily resolved. For comparison, the red dashed curves show the individual Lorentzian-shaped spectra for each of the three peaks in the absence of line-narrowing, and the black dashed curve their superposition, showing an asymmetry similar to that seen in the spontaneous Brillouin spectrum in Figure 2a (blue curve). This technique allows much more accurate measurements (within the cavity free spectral range of ~520 kHz) of the Brillouin frequency shifts than is possible from the spontaneous Brillouin spectrum, with a precision can be further improved by extending the fiber length.

Figure 6b shows the results for $[-2, +1]_P$ and $[+2, -1]_S$ modes in $C_6$ PCF. Three peaks at 10.994 GHz, 11.081 GHz and 11.126 GHz are easily resolved, showing good agreement with the simulations in Figure 2d. However, in the spontaneous Brillouin measurement in Figure 2a (red curve), a symmetric Lorentzian-shaped spectrum without side-peak was only observed. We attribute this to different interaction strengths in the spontaneous and lasing cases, the two additional Brillouin peaks being too weak to be spontaneously excited. We note in addition that line-narrowing in a Brillouin laser cavity can be potentially used in photoacoustic spectroscopy [21] and fibre sensing [22].

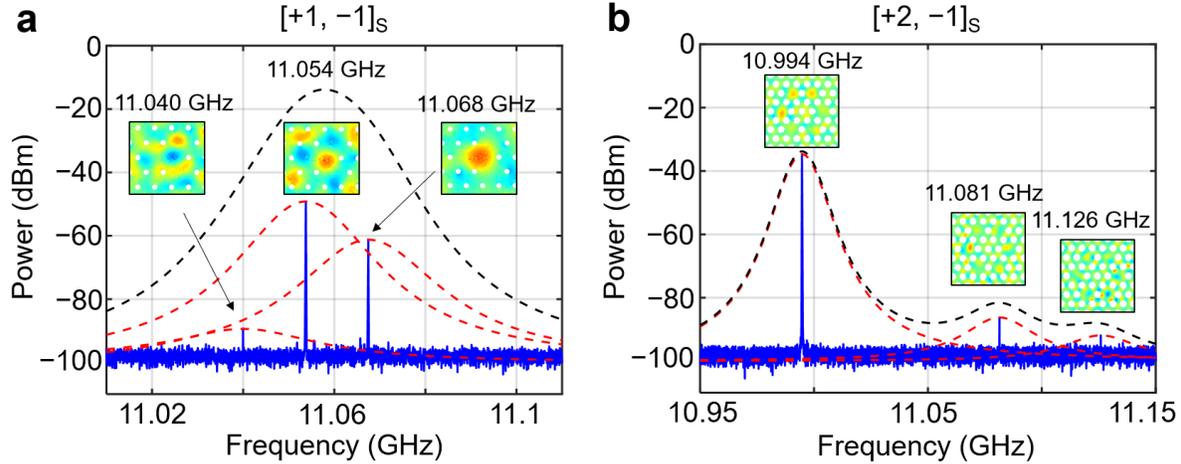

**Figure 6.** a) The spectrum of the $[+1, -1]_S$ lasing mode in $C_3$ PCF, measured by heterodyning the laser signal with a local oscillator at the pump frequency. The red dashed curves show the calculated Lorentzian functions for three Brillouin peaks and the black dashed curve their superposition. Insets, normalized axial displacements of three acoustic modes. b) same as a), but for lasing in the $[+2, -1]_S$ mode in $C_6$ PCF.

## 3. Conclusion

We constructed stable vortex Brillouin lasers using chiral three-fold and six-fold rotationally symmetric PCFs as gain media. Optoacoustic interactions between optical vortices and guided acoustic modes give rise to chiral stimulated Brillouin scattering in which power is transferred between pump and Stokes modes with opposite topological charge and spin. In contrast to other SBS-related vortex emission schemes, the vortex Brillouin laser in this paper can intrinsically maintain the vorticity of light in the whole laser cavity without need for mode control techniques. Laser line-narrowing permits the closely-spaced Brillouin peaks associated with different acoustic modes to be precisely resolved, in excellent agreement with theoretical predictions.

This is to our knowledge the first report of stable laser oscillation of optical vortices and acoustic modes in a Brillouin laser cavity. We note that annular core fibres [23], can stably preserve vortex modes and be used to construct the vortex Brillouin lasers. Vortex Brillouin lasers based on chiral PCF are of potential interest in optical tweezers, classic and quantum communications and quantum photon-phonon studies. High-resolution measurement of Brillouin peaks may be useful in acoustic spectroscopy and sensing.

**Acknowledgements:** This research is supported by the Max-Planck-Gesellschaft through an independent Max-Planck-Research Group.

**Competing interests:** Authors declare no competing interests.

**Data availability:** The data that support the plots within this paper and other findings of this study are available from the corresponding authors upon reasonable request.

## References

[1]  R. Y. Chiao, C. H. Townes, and B. P. Stoicheff, *Phys. Rev. Lett.* **1964**, 12, 592.

[2]  A. Kobyakov, M. Sauer, D. Chowdhury, *Adv. Opt. Photon.* **2010**, 2, 1.

[3]  B. J. Eggleton, C. G. Poulton, P. T. Rakich, Michael. J. Steel, G. Bahl, *Nat. Photon.* **2019**, 13, 664.

[4]  V. Lecoeuche, D. J. Webb, C. N. Pannell, D. A. Jackson, *Opt. Commun.* **1998**, 152, 263.

[5]  S. P. Smith, F. Zarinetchi, S. Ezekiel, *Opt. Lett.* **1991**, 16, 393.

[6]  J. Li, M.-G. Suh, K. Vahala, *Optica* **2017**, 4, 346.

[7]  M. Chen, M. Mazilu, Y. Arita, E. M. Wright, K. Dholakia, *Opt. Lett.* **2013**, 38, 4919.

[8]  N. Bozinovic, Y. Yue, Y. Ren, M. Tur, P. Kristensen, H. Huang, A. E. Willner, S. Ramachandran, *Science* **2013**, 340, 1545.

[9]  G. Vallone, V. D'Ambrosio, A. Sponselli, S. Slussarenko, L. Marrucci, F. Sciarrino, P. Villoresi, *Phys. Rev. Lett.* **2014**, 113, 060503.

[10] J. Wang, J. Zhang, A. Wang, X. Jiang, J. Yao, Q. Zhan, *Opt. Express* **2021**, 29, 18408.

[11] J. Xu, L. Zhang, X. Liu, L. Zhang, J. Lu, L. Wang, X. Zeng, *Opt. Lett.* **46**, 468.

[12] P. St.J. Russell, R. Beravat, G. K. L. Wong, *Phil. Trans. R. Soc. A* **2017**, 375, 20150440.

[13] X. Zeng, W. He, M. H. Frosz, A. Geilen, P. Roth, G. K. L. Wong, P. St.J. Russell, B. Stiller, Photon. Res. 2022, 10, 711.

[14] X. Zeng, P. St.J. Russell, C. Wolff, M. H. Frosz, G. K. L. Wong, B. Stiller, arXiv:2203.03680 [physics.optics], 7 Mar 2022.

[15] P. Roth, M. H. Frosz, L. Weise, P. St.J. Russell, G. K. L. Wong, *Opt. Lett.* **2021**, 46, 174.

[16] Y. Chen, P. St.J. Russell, *J. Opt. Soc. Am. B* **2021**, 38, 1173.

[17] V. Laude, A. Khelif, S. Benchbane, M. Wilm, T. Sylvestre, B. Kibler, A. Mussot, J. M. Dudley, H. Maillotte, *Phys. Rev. B* **2005**, 71, 045107.

[18] R. L. Fork and M. A. Pollack, *Phys. Rev.* **1965**, 139, A1408.

[19] B. Daino, P. Spano, M. Tamburrini, and S. Piazzolla, *IEEE J. Quantum Electron.* **1983**, 19, 266.

[20] L. Richter, H. Mandelberg, M. Kruger, P. McGrath, *IEEE J. Quantum Electron.* **1986**, 22, 2070.

[21] Qiang Wang, Zhen Wang, Jun Chang, Wei Ren, *Opt. Lett.* **2017**, 42, 2114.

[22] Joseph B. Murray, Alex Cerjan, Brandon Redding, *Optica* **2022**, 9, 80.

[23] P. Gregg, P. Kristensen, S. Ramachandran, *Optica* **2015**, 2, 267.

# Supplementary Information

## Optical vortex Brillouin laser

Xinglin Zeng[1], Philip St.J. Russell[1], Yang Chen[1], Zheqi Wang[1], Gordon K. L. Wong[1], Paul Roth[1], Michael H. Frosz[1] and Birgit Stiller[1,2]

1. Max-Planck Institute for the Science of Light, Staudtstr. 2, 91058 Erlangen, Germany
2. Department of Physics, Friedrich-Alexander-Universität, Staudtstr. 2, 91058 Erlangen, Germany

### S1. Heterodyne measurement setup

Light from a narrow linewidth (<1 kHz) 1550 nm continuous wave (CW) laser is split into pump and local oscillator (LO) signals at a a fibre coupler. The pump signal is then amplified in an EDFA and launched into the chiral PCF via an optical circulator. The circular polarization state is adjusted using a fibre polarization controller (FPC) placed before the circulator. The vortex generating module (polarizer, λ/4 plate and Q-plate) is optionally used to generate a circularly-polarized vortex-carrying pump signal. The noise-seeded Stokes signal from the PCF is delivered by the circulator and interferes with the LO signal using a second 90:10 fibre coupler. Narrow-band (6 GHz) notch filters placed in the path of the Stokes signal are used to filter out Fresnel reflections and Rayleigh scattering. The resulting beat-note is detected in the radio-frequency domain using a fast photodiode (PD) and the averaged Brillouin spectra is recorded with an electrical spectrum analyzer (ESA).

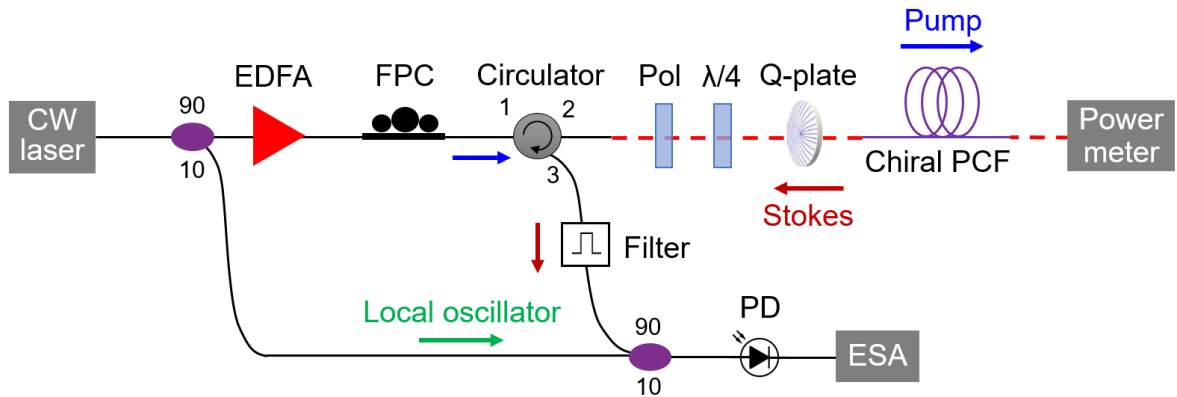

**Figure S1.** Heterodyne setup for the measurement of spontaneous Brillouin spectra. Pol: polarizer. λ/4: λ/4 plate.

## S2. Pump-seed setup

The pump-seed setup is shown in Figure S2. Both pump and seed were derived from a narrow linewidth 1550 nm CW laser, the seed light being frequency tuned using a single side-band modulator (SSBM). The pump signal was boosted by an EDFA and the polarization states of both pump and seed were controlled using FPCs. Vortex generating modules (circular polarizer, Q-plate and λ/2 plate) were optionally used to generate circularly-polarized vortex-carrying pump signals. After propagating backwards through the chiral PCF and interacting with pump signal, the seed signal is reflected by a beam splitter (BS), filtered by a circular polarizer (polarizer and λ/4 plate) and finally delivered to NBA system for gain coefficient measurement.

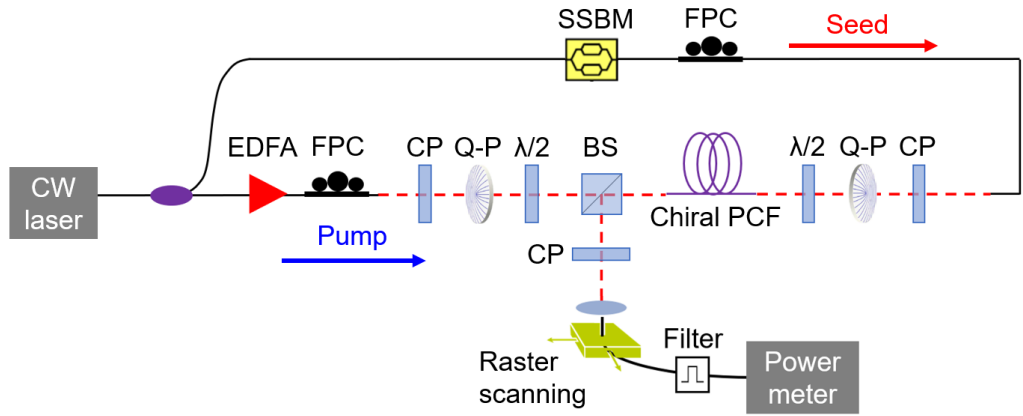

**Figure S2.** Pump-seed experimental setup for gain coefficient measurement. CP: circular polarizer (polarizer and λ/4 plate). Q-P: Q-plate. λ/2: λ/2 plate.